\documentclass{phostproc}

\usepackage{bm}

\title{Comprehensive stellar seismic analysis : A preliminary application of Whosglad to 16 Cygni system.}
\author{M.~Farnir$^{1}$,
 M-A.~Dupret$^{1}$,
 S.J.A.J.~Salmon$^{1}$,
 A. Noels$^{1}$,
 G. Buldgen$^{2}$}

\affiliation{$^{1}$ Institut d’Astrophysique et Géophysique de l’Université de Liège, Allée du 6 août 17, 4000 Liège, Belgium \\
			 $^{2}$ School of Physics and Astronomy, University of Birmingham, Edgbaston, Birmingham B15 2TT, UK}

\shorttitle{Comprehensive stellar seismic analysis}
\shortauthors{M.~Farnir \textit{et al.}}

\abs{We present a first application of Whosglad method to the components A and B of the 16 Cygni system. The method was developed to provide a comprehensive analysis of stellar oscillation spectra. It defines new seismic indicators which are as uncorrelated and precise as possible and hold detailed information about stellar interiors. Such indicators, as illustrated in the present paper, may be used to generate stellar models via forward seismic modeling. Finally, seismic constraints retrieved by the method provide realistic stellar parameters.}

\begin{document}

\maketitle

\section{Introduction}
As the quality of asteroseismic data improves, due to space missions such as CoRoT \citep{2009IAUS..253...71B} and \emph{Kepler} \citep{2010AAS...21510101B}, it becomes possible to study the frequency signature of acoustic glitches. Acoustic glitches are oscillating features visible in oscillation spectra due to a sharp variation in the stellar structure. It was first \citet{1988IAUS..123..151V} who showed that such a sharp variation can be directly observed in the frequencies. Subsequently, \citet{1990LNP...367..283G} showed that it is also visible in the second frequency differences. Therefore, the detection and study of acoustic glitches may provide essential information to better constraint and understand the stellar structure. For example, constraints about the localisation of the base of the envelope convective zone or the surface helium abundance may be derived. This has already been the subject of several studies (\citealt{2000MNRAS.316..165M,2004MNRAS.350..277B,2014ApJ...790..138V}, to name a few). However, it is of prime importance to take advantage of as much of the information available in the oscillation spectrum as possible. Furthermore, the information has to be treated in a statistically relevant way. For these reasons, we developed a new method to analyse the whole oscillation spectrum (both smooth and oscillating -- glitch -- part) simultaneously in a homogeneous and statistically relevant way. Using our method, we define new seismic indicators which are as uncorrelated to each other as possible. Those indicators may then be used as constraints for forward seismic modeling to provide improved models and the proper correlations between the resulting parameters for a given target. 

These proceedings will focus on examples of adjustments of 16 Cygni A and B using our method. Therefore, we only present what is necessary to the adjustments. The present paper is organised as follows: Sect.~\ref{Sec:Met} presents the aspects of the method necessary to understand the adjustments. Then, we define the indicators used in the calibrations in Sect.~\ref{Sec:Ind}. In Sect.~\ref{Sec:Fit} we present our adjustments of 16 Cygni A and B. Finally, a small conclusion to the paper is given in Sect.~\ref{Sec:Con}.

\section{Method}\label{Sec:Met}
The present section constitutes a synthesised version of Sect.~2 in \citet{2019A&A...622A..98F} to provide the reader with only the necessary material to understand how the adjustments were realised.

The developed method, \emph{WhoSGlAd} -- for \textbf{Who}le \textbf{S}pectrum and \textbf{Gl}itches \textbf{Ad}justment --, relies on linear algebra and \emph{Gram-Schmidt}'s \citep{Gram,Schmidt} algorithm to provide independent seismic indicators that may then be used in the framework of stellar modeling. The use of \emph{Gram-Schmidt}'s algorithm ensures the independence of the different fitted coefficients. Then, those are combined to build seismic indicators that are as uncorrelated to one another as possible. Their covariances are properly computed.

We consider a Euclidean vector space that is the set of $N$ observed oscillation frequencies $\nu_i$. The standard deviation for each frequency is written $\sigma_i$. Given two vectors $\mathbf{x}$ and $\mathbf{y}$, we may define their scalar product : 
\begin{equation}\label{Eq:ScaPro}
\left\langle \mathbf{x} | \mathbf{y} \right\rangle~=~\sum\limits^N_{i=1} \frac{x_i y_i}{\sigma^2_i}.
\end{equation}
This definition may now be used to express the merit function that will be used to compare two sets of frequencies (e.g. theoretical and observed frequencies):
\begin{equation}\label{Eq:Chi}
\chi^2~=~\Vert\bm{\nu}_\textrm{obs}-\bm{\nu}_\textrm{th}\Vert^2~=~\sum\limits^N_{i=1} \frac{\left(\nu_{\textrm{obs},i}-\nu_{\textrm{th},i}\right)^2}{\sigma^2_i},
\end{equation} 
with $\bm{\nu}_\textrm{th}$ and $\bm{\nu}_\textrm{obs}$, the theoretical and observed\footnote{We denote by the subscript \textit{obs} both the observed frequencies and the frequencies derived from a reference model -- which constitutes an artificial target -- and we denote by the subscript \textit{th} the frequencies that we adjust to those observations.} frequencies. In practice, it is convenient  to label frequencies by their radial order $n$ and spherical degree $l$. thus, from now on, we will use this notation rather than the index $i$.

In the presence of a glitch, \citet{2007MNRAS.375..861H} showed that the oscillatory component in frequencies due to the glitch can be isolated from the rest of the spectrum, called the smooth component. Thus, to represent observed frequencies, we define a vector subspace that is typically a polynomial space -- the smooth component -- associated with an oscillating component -- the glitch --. This is very similar to what has been done by \citet{2014ApJ...790..138V}.

The method consists in the projection of the observed and theoretical frequencies over the vector subspace. Then, we define seismic indicators from the projections. Some of them are defined in Sect.~\ref{Sec:Ind}.
To do so, it is useful to define an orthonormal basis over the vector subspace. This is done via \emph{Gram-Schmidt}'s orthogonalisation process associated with the definition of the scalar product (\ref{Eq:ScaPro}).

If we write $j$ and $j_0$ the indices associated with the basis elements, $\mathbf{p}_{j}$ the former basis elements, $\mathbf{q}_{j_0}$ the orthonormal basis elements, and $R^{-1}_{j,j_0}$ the transformation matrix, we may relate the old basis elements with the new orthonormalised ones via the relation:
\begin{equation}\label{Eq:Bas}
q_{j_0}(n,l)~=~\sum\limits_{j\leq j_0}R^{-1}_{j,j_0}p_{j}(n,l),
\end{equation} 
where the dependence in $n$ and $l$ translates that the basis elements are evaluated at each observed value of the radial order $n$ and the spherical degree $l$. This is essential to note as the set of observed frequencies and standard deviations will be different for each value of $l$. Therefore, the basis elements will be different for each of them.

A crucial point of the method relies on the fact that the frequencies will be projected on the basis in a specific order. This will allow to produce the lowest possible $\chi^2$ value but also will be of prime importance to build the indicators. The projections of the frequencies on the basis will be written $a_{j}~=~ \left\langle \bm{\nu} \vert \mathbf{q}_{j} \right\rangle$ and the adjusted frequencies become : 
\begin{equation}\label{Eq:FreFit}
\nu_{f}(n,l)~=~\sum\limits_j a_{j} q_{j}(n,l).
\end{equation}
One should be reminded that the orthogonalisation ensures that the coefficients $a_j$ are independent of each other. In other words, their covariance matrix is the identity matrix and their standard deviation becomes equal to $1$.

To describe the smooth part of the spectrum, we use the following succession of polynomials:

\begin{equation}\label{Eq:PolSmo}
p_{lk}(n,l')~=~\delta_{ll'} p_{k}(n),
\end{equation}
where $p_k(n)~=~n^k$, $k~=~0,1~\textrm{or}~2$ and $\delta_{ll'}$ is the \emph{Kronecker} delta which compares two spherical degrees $l$ and $l'$.

Then, to represent the contribution of the glitch to the frequencies, we used a linearised form based on the one used by \citet{2014ApJ...790..138V}. The detailed formulation we used to describe the glitch is not needed in the present paper. Therefore, it is not given here.

\section{Indicators}\label{Sec:Ind}
We present in this section the indicators used in Sect.~\ref{Sec:Fit}. An in depth study of those indicators as well as some other useful ones is presented in \citet{2019A&A...622A..98F}.

\subsection{Large separation}\label{Sec:LSep}
The first seismic indicator that comes to mind is the large separation which holds a local (i.e. based on the individual frequencies) and an asymptotic definition. Thus, we defined an estimator for the large separation in our formalism. To do so, we take inspiration in the asymptotic approximation. In the high frequency regime ($n \gg l$), the asymptotic approximation is valid and the frequencies may be expressed, to the first order, as \citep{1986HiA.....7..283G}:
\begin{equation}\label{Eq:AsyFre}
\nu\left(n,l\right)~\simeq~\left( n+\frac{l}{2}+\epsilon \right)\Delta\nu,
\end{equation}
where $\Delta\nu~=~\left(2\int^{R_*}_0 \frac{dr}{c(r)} \right)^{-1}$ is the asymptotic large frequency separation, $c(r)$ is the adiabatic sound speed, and $R_*$ is the radius at the photosphere of the star. 
We note that, in such a formulation, the large separation is the slope in $n$ of the straight line fitting at best the frequencies. Then, because we first project the frequencies on the zero order in $n$ polynomial, we adjust the frequencies to a constant value for each spherical degree, its mean value $\overline{\nu}_l$. Finally, the frequencies are projected on a first order polynomial and the corresponding coefficient indeed corresponds to the slope of the best-fit straight line. Therefore, it constitutes an estimator of the large separation and takes the form:
\begin{equation}\label{Eq:Dnul}
\Delta_l~=~a_{l,1} R^{-1}_{l,1,1}.
\end{equation}
One should note that the previously defined indicator depends on the spherical degree. Thus, we may average those and retrieve the mean large separation. Knowing that the standard deviation of $a_{l,1}$ is $1$, $(R^{-1}_{l,1,1})^2$ is the variance of $\Delta_l$ we have: 
\begin{equation}\label{Eq:Dnu}
\Delta~=~\frac{\sum\limits_l a_{l,1}/R^{-1}_{l,1,1}}{\sum\limits_l 1/(R^{-1}_{l,1,1})^2}.
\end{equation}

\subsection{Small separation ratios}\label{Sec:SSep}
\citet{2003A&A...411..215R} showed that dividing the the small separations $d_{01}(n)$ and $d_{02}(n)$ by the large separation allowed to minimise the impact of the surface effects on the resulting indicators. Those may be defined locally as:
\begin{equation}\label{Eq:r01R}
r_{01}(n)~=~\frac{d_{01}(n)}{\Delta\nu_1(n)}~=~\frac{\nu(n - 1, 1) - 2\nu(n, 0) + \nu(n, 1)}{2(\nu(n,1) - \nu(n-1,1))},
\end{equation}
\begin{equation}\label{Eq:r02R}
r_{02}(n)~=~\frac{d_{01}(n)}{\Delta\nu_1(n)}~=~\frac{\nu(n,0) - \nu(n-1, 2)}{\nu(n,1) - \nu(n-1,1)}.
\end{equation}
The ratio $r_{01}(n)$ and $r_{02}(n)$ represent the local spacing (i.e. at a given radial order) between the ridges of degrees $0$ and $1$, and $0$ and $2$ respectively in an \emph{Échelle} diagram \citep{1983SoPh...82...55G}.
If we now take interest in the mean spacing between the ridges of degrees $0$ and $l$ it corresponds to $\left(\overline{\nu_0}-\overline{\nu_l}\right)/\Delta_0$. Then assuming Eq.~\ref{Eq:AsyFre} to be exact, it is equal to $\overline{n_0} + \epsilon_0 - (\overline{n_l} + \epsilon_l + l/2)$. Finally, adding $-\overline{n_0} + \overline{n_l} + l/2$ to the previous expression approaches it to $\epsilon_0 - \epsilon_l$. This leads to the following expression for the small separation ratios estimators:
\begin{equation}
\hat{r}_{0l}~=~\frac{\overline{\nu_0}-\overline{\nu_l}}{\Delta_0}+\overline{n_l}-\overline{n_0}+\frac{l}{2}.\label{Eq:r0l}
\end{equation}

It has been shown by \citet{2019A&A...622A..98F} that such indicators are indeed almost insensitive to surface effects.

\subsection{Helium glitch amplitude}\label{Sec:AHe}
To provide an indicator of the helium glitch amplitude, which is thought to be closely related to the surface helium content \citep{2004MNRAS.350..277B,2007MNRAS.375..861H,2014ApJ...790..138V,2019A&A...622A..98F}, we calculated the norm of the helium glitch component. This yields:
\begin{equation}
A_\textrm{He}~=~\Vert \delta\nu_{g,\textrm{He}} \Vert~=~\sqrt{\sum\limits_j a_{j,\textrm{He}}^2},
\end{equation}
where $\delta\nu_{g,\textrm{He}}$ is the fitted helium glitch and $a_{j,\textrm{He}}$ represents the several coefficients fitted to the glitch. \citet{2019A&A...622A..98F} have shown that this indicator is indeed correlated with the surface helium content. However, attention has to be paid because, as noted by \citet{2004MNRAS.350..277B}, the helium glitch amplitude is anti-correlated to metallicity. This is the case for our indicator as well. Nevertheless, a simple toy model of the first adiabatic exponent may be used to understand (in a qualitative way) those correlations.

\section{Adjustments}\label{Sec:Fit}
In this section, we present several preliminary adjustments of models (in the framework of direct seismic modeling) to 16 Cygni A and B observed seismic indicators determined solely using our method Whosglad. We used \emph{Levenberg-Marquardt}'s algorithm to fit theoretical values to the set of observed seismic indicators shown in Table \ref{Tab:SeiInd}. The free parameters were the mass $M$, the age $t$, the initial central hydrogen abundance $X_0$ and the initial ratio of hydrogen over metals abundances $Z/X_0$. The seismic indicators, used as constraints, have been computed using the frequencies calculated by \citet{2015MNRAS.446.2959D} and corrected for the surface effects using \citet{2008ApJ...683L.175K}'s prescription of which the coefficients $a$ and $b$ have been calibrated by \citet{2015A&A...583A.112S}.
We realised two independent calibrations, using either the metal mixture of AGSS09 \citep{2009ARA&A..47..481A} or that of GN93 \citep{1993oee..conf...15G}, for each of the components A and B of the 16 Cygni system. The models were calculated using the CLES \citep{2008Ap&SS.316...83S} stellar evolution code and the theoretical frequencies were computed with the LOSC \citep{2008Ap&SS.316..149S} oscillation code. The models used the FreeEOS software \citep{2003ApJ...588..862C} to generate the equation of state table, the reaction rates prescribed by \citet{2011RvMP...83..195A} and the OPAL opacity table \citep{1996ApJ...464..943I} combined with that of \citet{2005ApJ...623..585F} at low temperatures. Moreover, the mixing inside convective regions was computed according to the mixing length theory \citep{1968pss..book.....C} and using the value $\alpha_\textrm{MLT}~=~l/H_p=~1.82$ (where $l$ is the mixing length and $H_p$ the pressure scale height) that we obtained via a solar calibration. Microscopic diffusion was taken into account in the computation by using \citet{1994ApJ...421..828T}'s routine. For each model, the temperature at the photosphere and the conditions above the photosphere are determined by using an Eddington $T(\tau)$ relationship.

\begin{figure}
\centering
\includegraphics[width=0.85\linewidth]{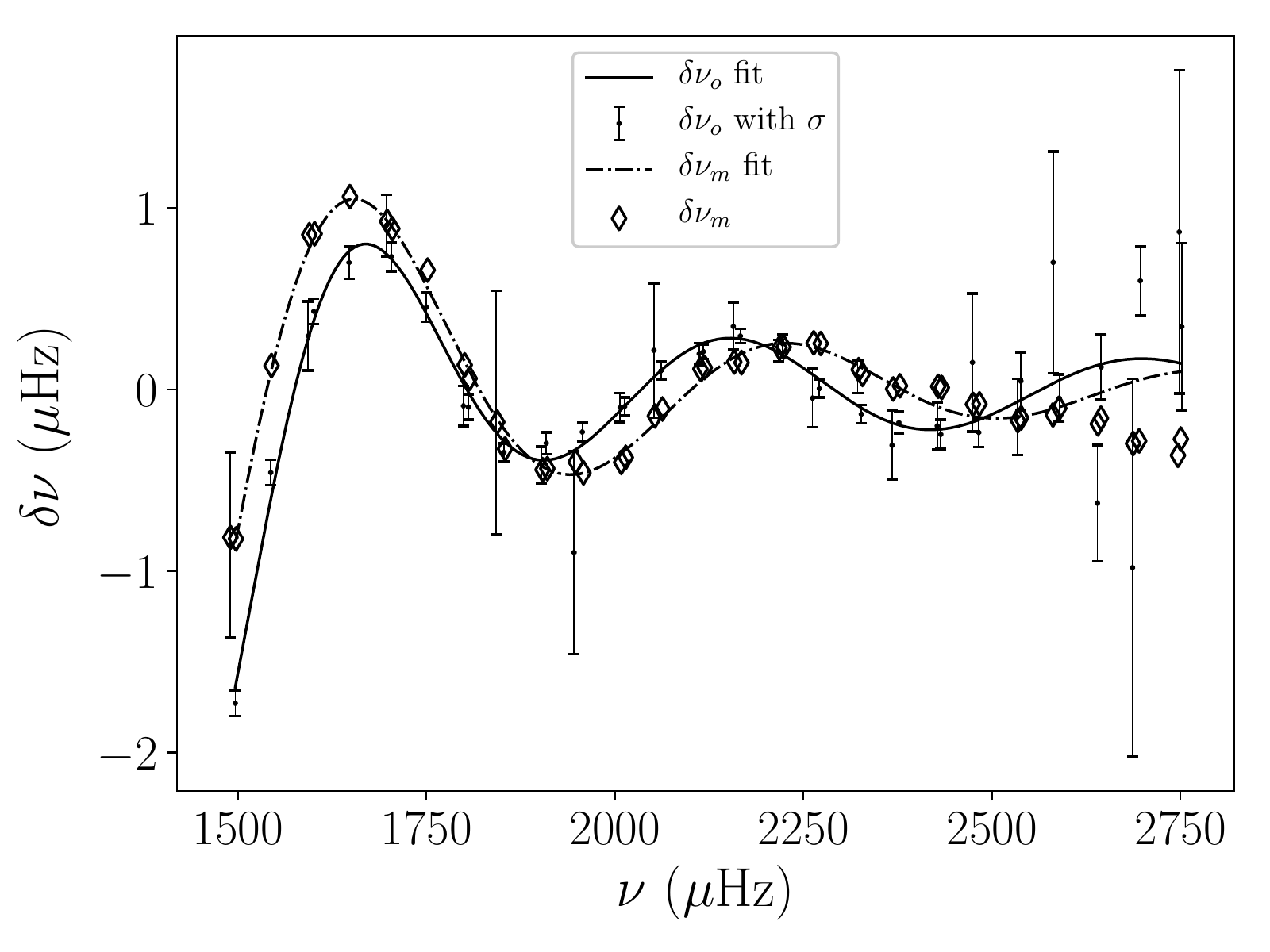}
\caption{16 Cygni A observed and best fit model fitted glitches (solid and dashed line respectively). Observed frequencies and their incertitudes are represented as errorbars while best fit frequencies are symbolised by diamonds.}\label{Fig:FitA}
\end{figure}

\begin{figure}
\centering
\includegraphics[width=0.85\linewidth]{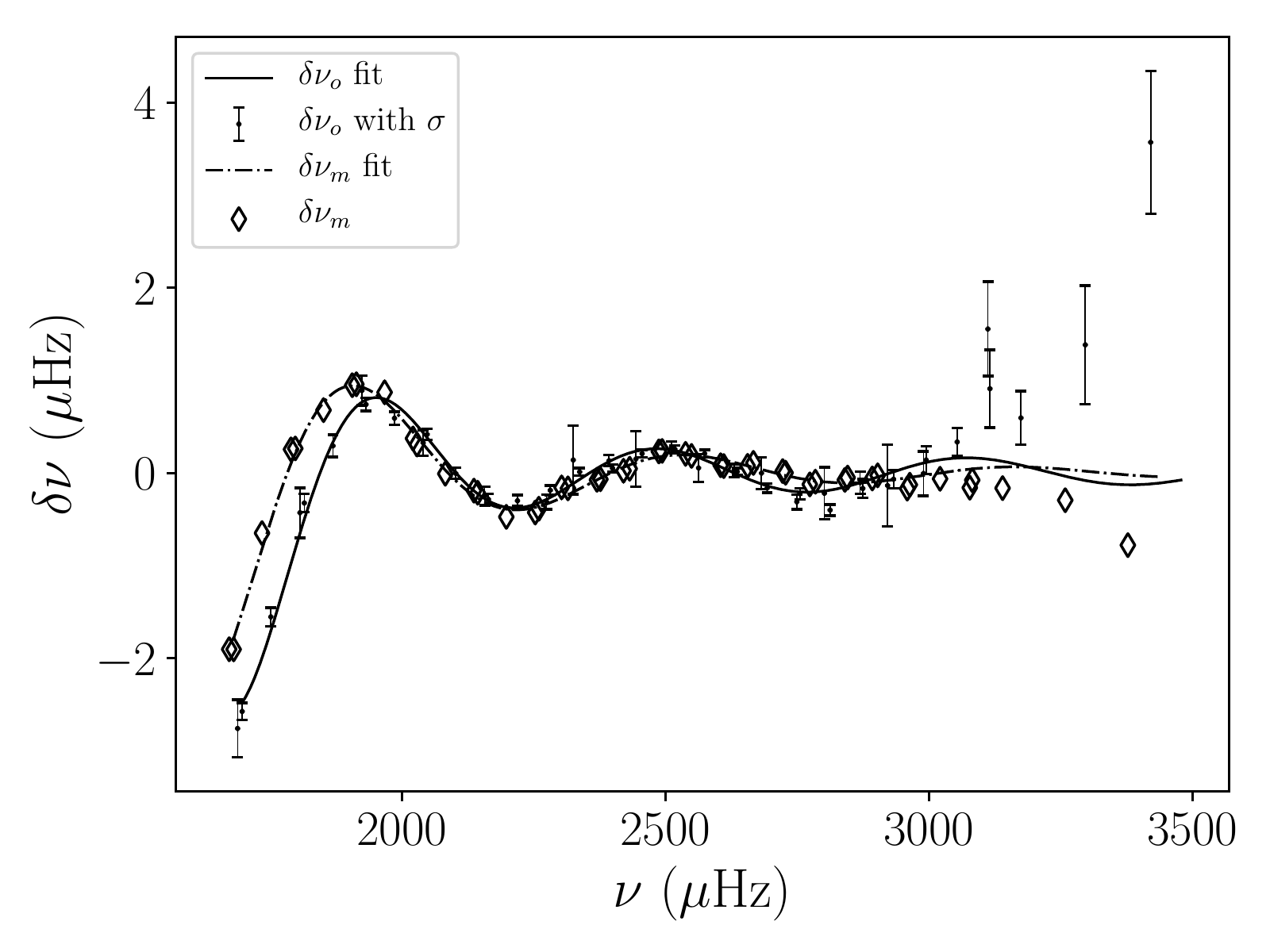}
\caption{16 Cygni B observed and best fit model fitted glitches (solid and dashed line respectively). Observed frequencies and their incertitudes are represented as errorbars while best fit frequencies are symbolised by diamonds.}\label{Fig:FitB}
\end{figure}

Figs.~\ref{Fig:FitA} and \ref{Fig:FitB} provide comparisons between the observed and best fit model glitches for both component of the system (A and B respectively). For component A, we observe that the agreement between both glitches is good, as expected from the amplitude adjustment. However, we note that there is an offset in phase. To provide better results, the phase might then be added to the constraints. Nevertheless, it is still not clear that this maneuver will improve the results in a significant way and this should be tested in future studies. Next, for component B, we also note an offset in phase. Moreover, we see that the highest frequencies (compared to $\nu_\textrm{max}$) are not well represented by the glitch function. Such a discrepancy should come from the surface effects and the prescription chosen to account for them as their influence is greater for the high frequencies. In addition, this is of little influence on the quality of the results as the glitch is of greater amplitude in the low frequency regime, as opposed to the surface effects correction.

\begin{figure}
\centering
\includegraphics[width=0.85\linewidth]{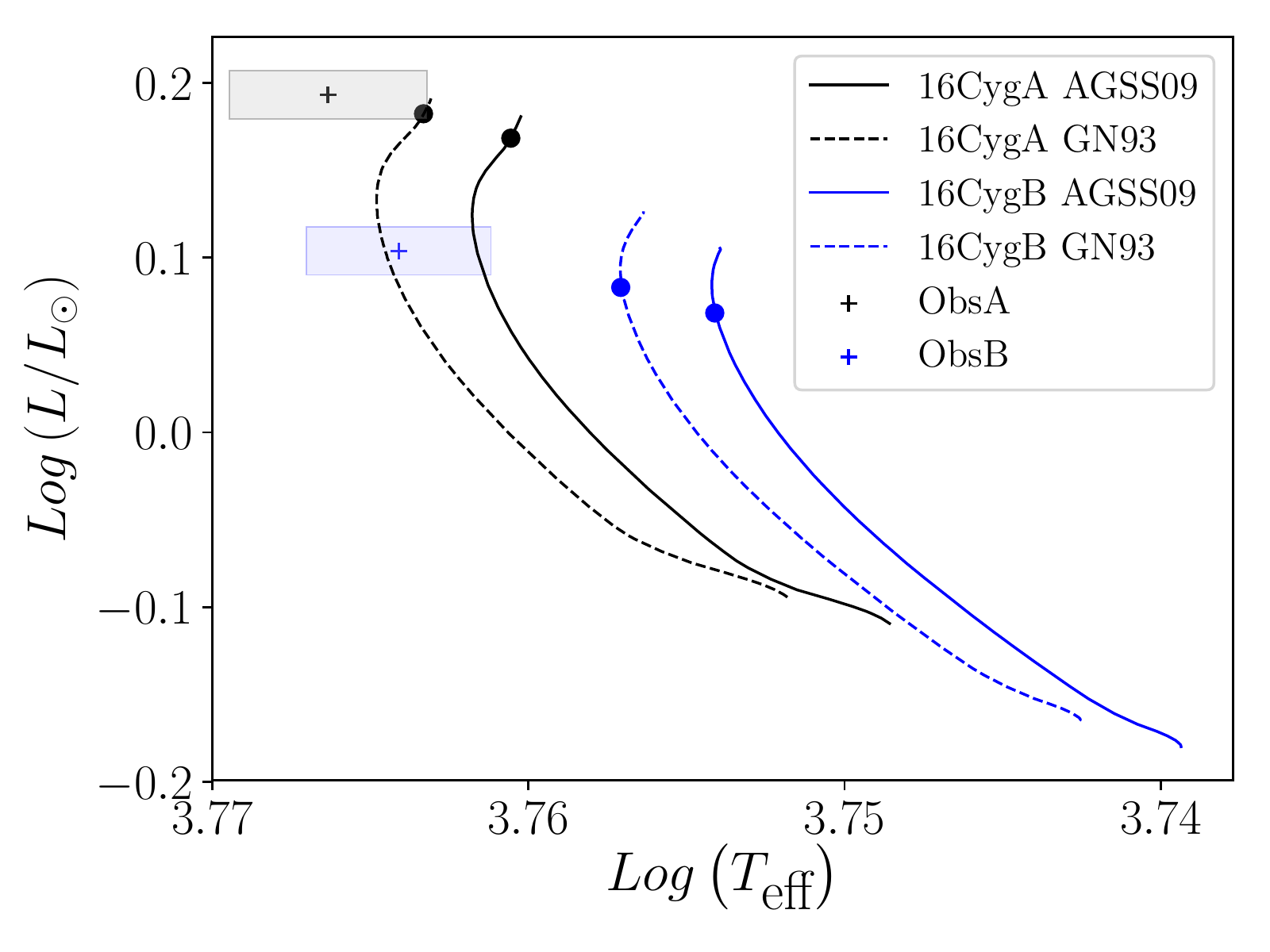}
\caption{Best fit model tracks in the HR diagram. Black tracks correspond to 16 Cygni A while the blue ones correspond to the B component. The line style symbolises the solar mixture: straight lines are for AGSS09 and dashed lines are for GN93. The best fit model for each trach is represented as a dot. Observed values for the luminosities and effective temperatures as well as their respective incertitudes are shown as a box.}\label{Fig:Tra}
\end{figure}

Tables \ref{Tab:ResA} and \ref{Tab:ResB} show the results obtained. The presented standard deviations are solely the ones intrinsic to the method. We observe that for both components of the system, using the GN93 solar mixture rather than the AGSS09 one produces similar changes in the calibrated stellar parameter, both in direction and magnitude. Moreover, it is interesting to note that, considering one solar mixture at a time, the derived ages for both component lie within one standard deviation of one another even though both adjustments have been realised independently without imposing a common value.
We also note that the values we retrieve for the surface helium abundance are in perfect agreement with \citet{2014ApJ...790..138V} who calculated values lying in $Y_f~\in~[0.23,0.25]$ and $[0.218,0.260]$ for A and B respectively. Moreover, we observe that these calibrations tend to favour the GN93 abundances as the computed effective temperatures and luminosities lie closer to the observed ones (respectively $5839~\pm~42 K$ and $5809~\pm~39 K$ \citet{2013MNRAS.433.1262W} and $1.56~\pm~0.05$ and $1.27~\pm~0.04$ \citet{2012ApJ...748L..10M} for 16 Cyg A and B). This is visible in Fig.~\ref{Fig:Tra} where the evolutionary tracks of the best fit models are represented in a HR diagram. The dots symbolise the best fit models. The constraints in luminosity and temperature are represented as boxes. As one may observe, those constraints are almost satisfied in the case of 16 Cygni A with the GN93 abundances. To reconcile the other models with the observations, one may free the mixing length parameter $\alpha_\textrm{MLT}$. Finally, one should be careful when considering those results as many different physics should still be tested and non-seismic constraints (for example the metallicity which is more than twice bigger than the measurements of \citet{2009A&A...508L..17R} for both components) should be included in thr constraints to provide a thorough analysis of the 16 Cygni system.

\begin{table}
\centering
\begin{tabular}{ccccc}
\hline
 & \multicolumn{2}{c}{16 Cyg A} & \multicolumn{2}{c}{16 Cyg B} \\
Indicator & Value & $\sigma$ & Value & $\sigma$ \\ 
\hline
\hline\\[-0.8em]
$\Delta (\mu Hz)$ & $104.088$ & $0.005$ & $118.614$ & $0.004$ \\
$A_\textrm{He}$ & $30.4$ & $1.0$ & $29.8$ & $1.0$ \\
$\hat{r}_{01}$ & $0.0362$ & $0.0002$ & $0.0251$ & $0.0002$ \\
$\hat{r}_{02}$ & $0.0575$ & $0.0003$ & $0.0555$ & $0.0003$ \\
\hline 
\end{tabular}
\caption{Observed seismic indicators.}\label{Tab:SeiInd}
\end{table}

\begin{table}
\centering
\begin{tabular}{ccccc}
\hline
 & \multicolumn{2}{c}{AGSS09} & \multicolumn{2}{c}{GN93} \\
Quantity & Value & $\sigma$ & Value & $\sigma$ \\ 
\hline
\hline\\[-0.8em]
$M~(M_\odot)$ & $1.057$ & $0.021$ & $1.067$ & $0.011$ \\
$t~(\textrm{Gyr})$ & $6.8$ & $0.2$ & $6.6$ & $0.1$ \\
$(Z/X)_0$ & $0.035$ & $0.002$ & $0.039$ & $0.002$ \\
$X_0$ & $0.68$ & $0.01$ & $0.69$ & $0.01$ \\
%$Y_0$ & $0.29$ & $0.02$ & $0.28$ & $0.01$ \\
$Y_f$ & $0.242$ & $0.016$ & $0.237$ & $0.010$ \\
$[Fe/H]$ & $0.19$ & $0.03$ & $0.24$ & $0.03$ \\
$R~(R_\odot)$ & $1.22$ & $/$ & $1.22$ & $/$ \\
$L~(L_\odot)$ & $1.47$ & $/$ & $1.52$ & $/$ \\
$T_\textrm{eff}~(K)$ & $5762$ & $/$ & $5798$ & $/$ \\
\hline 
\end{tabular}
\caption{Fitted 16 Cyg A parameters.}\label{Tab:ResA}
\end{table}

\begin{table}
\centering
\begin{tabular}{ccccc}
\hline
 & \multicolumn{2}{c}{AGSS09} & \multicolumn{2}{c}{GN93} \\
Quantity & Value & $\sigma$ & Value & $\sigma$ \\ 
\hline
\hline\\[-0.8em]
$M~(M_\odot)$ & $1.028$ & $0.009$ & $1.037$ & $0.012$ \\
$t~(\textrm{Gyr})$ & $7.07$ & $0.06$ & $6.89$ & $0.12$ \\
$(Z/X)_0$ & $0.033$ & $0.002$ & $0.037$ & $0.002$ \\
$X_0$ & $0.696$ & $0.008$ & $0.701$ & $0.008$ \\
%$Y_0$ & $0.281$ & $0.008$ & $0.273$ & $0.009$ \\
$Y_f$ & $0.237$ & $0.008$ & $0.231$ & $0.009$ \\
$[Fe/H]$ & $0.17$ & $0.03$ & $0.22$ & $0.03$ \\
$R~(R_\odot)$ & $1.12$ & $/$ & $1.12$ & $/$ \\
$L~(L_\odot)$ & $1.17$ & $/$ & $1.21$ & $/$ \\
$T_\textrm{eff}~(K)$ & $5676$ & $/$ & $5715$ & $/$ \\
\hline 
\end{tabular}
\caption{Fitted 16 Cyg B parameters.}\label{Tab:ResB}
\end{table}

\section{Conclusion}\label{Sec:Con}
We developed a new method to analyse the oscillation spectrum as a whole -- both smooth and glitch part are treated simultaneously. This allows to provide fitted parameters which are completely independent of each other (thanks to Gram-Schmidt's procedure). Seismic indicators are then built from these coefficients in such a way that they are as uncorrelated as possible. This allows to use those indicators in seismic modeling while being able to know the proper correlation between the fitted constraints.
Finally, we demonstrated the aptitude of the method to provide proper constraints by analysing the oscillation spectra of 16 Cygni A and B and calculating best-fit models in the framework of forward seismic modeling. A more detailed analysis of both 16 Cygni A and B should be realised in a future paper.

\section*{Acknowledgments}
M.F. is supported by the FRIA (Fond pour la Recherche en Industrie et Agriculture) - FNRS PhD grant.\\
S.J.A.J.S. is funded by ARC grant for Concerted Research Actions, financed by the Wallonia-Brussels Federation. \\
GB acknowledges support from the ERC Consolidator Grant funding scheme ({\em project ASTEROCHRONOMETRY}, G.A. n. 772293

\bibliographystyle{phostproc}
\bibliography{bibli}

\end{document}